\documentclass[a4paper,11pt]{article}
\usepackage{jinstpub} 
\usepackage{lineno}
\usepackage{subcaption}

\title{\boldmath Muon Tagging with Flash ADC Waveform Baselines}

\author[a]{D. H. Lee,} 
\author[b]{M. K. Cheoun,} 
\author[c]{J. H. Choi,} 
\author[d]{J. Y. Choi,} 
\author[e,h]{T. Dodo,} 
\author[f]{J. Goh,} 
\author[g]{M. Harada,} 
\author[h,g]{S. Hasegawa,} 
\author[f]{W. Hwang,} 
\author[i]{T. Iida,} 
\author[j]{H. I. Jang,} 
\author[k]{J. S. Jang,} 
\author[d]{K. K. Joo,} 
\author[l]{D. E. Jung,} 
\author[m]{S. K. Kang,} 
\author[g]{Y. Kasugai,} 
\author[n]{T. Kawasaki,} 
\author[d]{E. M. Kim,} 
\author[o]{E. J. Kim,} 
\author[p]{S. B. Kim,} 
\author[d]{S. Y. Kim,} 
\author[g]{H. Kinoshita,} 
\author[n]{T. Konno,} 
\author[q]{C. Little,} 
\author[a]{T. Maruyama,} 
\author[q]{E. Marzec,} 
\author[g]{S. Masuda,} 
\author[g]{S. Meigo,} 
\author[d]{D. H. Moon,} 
\author[r]{T. Nakano,} 
\author[s]{M. Niiyama,} 
\author[a]{K. Nishikawa,}  
\author[c]{M. Y. Pac,} 
\author[t]{B. J. Park,} 
\author[d]{H. W. Park,} 
\author[b]{J. B. Park,} 
\author[d]{Jisu Park,} 
\author[t]{J. S. Park,} 
\author[d]{R. G. Park,} 
\author[u]{S. J. M. Peeters,} 
\author[v,l]{C. Rott,} 
\author[t]{J. W. Ryu,} 
\author[g]{K. Sakai,} 
\author[g]{S. Sakamoto,} 
\author[r]{T. Shima,} 
\author[a]{C. D. Shin, \footnote{Corresponding author.}} 
\author[q]{J. Spitz,} 
\author[e]{F. Suekane,} 
\author[r]{Y. Sugaya,} 
\author[g]{K. Suzuya,} 
\author[i]{Y. Takeuchi,} 
\author[p]{W. Wang,} 
\author[p]{W. Wei,} 
\author[g]{Y. Yamaguchi,} 
\author[w]{M. Yeh,} 
\author[c]{I. S. Yeo,} 
\author[l]{and I. Yu} 

\affiliation[a]{High Energy Accelerator Research Organization (KEK),\\ 1-1 Oho, Tsukuba, Ibaraki, 305-0801, Japan}
\affiliation[b]{Department of Physics, Soongsil University,\\ 369 Sangdo-ro, Dongjak-gu, Seoul, 06978, Korea}
\affiliation[c]{Laboratory for High Energy Physics, Dongshin University,\\ 67, Dongshindae-gil, Naju-si, Jeollanam-do, 58245, Korea}
\affiliation[d]{Department of Physics, Chonnam National University,\\ 77, Yongbong-ro, Buk-gu, Gwangju, 61186, Korea}
\affiliation[e]{Research Center for Neutrino Science, Tohoku University,\\ 6-3 Azaaoba, Aramaki, Aoba-ku, Sendai 980-8578, Japan}
\affiliation[f]{Department of Physics, Kyung Hee University,\\ 26, Kyungheedae-ro, Dongdaemun-gu, Seoul 02447, Korea}
\affiliation[g]{J-PARC Center, JAEA,\\ 2-4 Shirakata, Tokai-mura, Naka-gun, Ibaraki 319-1195, Japan}
\affiliation[h]{Advanced Science Research Center, JAEA,\\ 2-4 Shirakata, Tokai-mura, Naka-gun, Ibaraki 319-1195, Japan}
\affiliation[i]{Faculty of Pure and Applied Sciences, University of Tsukuba,\\ Tennodai 1-1-1, Tsukuba, Ibaraki, 305-8571, Japan}
\affiliation[j]{Department of Fire Safety, Seoyeong University,\\ 1 Seogang-ro, Buk-gu, Gwangju, 61268, Korea}
\affiliation[k]{GIST College, Gwangju Institute of Science and Technology,\\ 123 Cheomdangwagi-ro, Buk-gu, Gwangju, 61005, Korea}
\affiliation[l]{Department of Physics, Sungkyunkwan University,\\ 2066, Seobu-ro, Jangan-gu, Suwon-si, Gyeonggi-do, 16419, Korea}
\affiliation[m]{School of Liberal Arts, Seoul National University of Science and Technology,\\ 232 Gongneung-ro, Nowon-gu, Seoul, 139-743, Korea}
\affiliation[n]{Department of Physics, Kitasato University,\\ 1 Chome-15-1 Kitazato, Minami Ward, Sagamihara, Kanagawa, 252-0329, Japan}
\affiliation[o]{Division of Science Education, Jeonbuk National University, \\ 567 Baekje-daero, Deokjin-gu, Jeonju-si, Jeollabuk-do, 54896, Korea}
\affiliation[p]{School of Physics, Sun Yat-sen (Zhongshan) University,\\ Haizhu District, Guangzhou, 510275, China}
\affiliation[q]{University of Michigan,\\ 500 S. State Street, Ann Arbor, MI 48109, U.S.A.}
\affiliation[r]{Research Center for Nuclear Physics, Osaka University,\\ 10-1 Mihogaoka, Ibaraki, Osaka, 567-0047, Japan}
\affiliation[s]{Department of Physics, Kyoto Sangyo University,\\ Motoyama, Kamigamo, Kita-Ku, Kyoto-City, 603-8555, Japan}
\affiliation[t]{Department of Physics, Kyungpook National University,\\ 80 Daehak-ro, Buk-gu, Daegu, 41566, Korea}
\affiliation[u]{Department of Physics and Astronomy, University of Sussex,\\ Falmer, Brighton, BN1 9RH, U.K.}
\affiliation[v]{Department of Physics and Astronomy, University of Utah,\\ 201 Presidents' Cir, Salt Lake City, UT 84112, U.S.A}
\affiliation[w]{Brookhaven National Laboratory,\\ Upton, NY 11973-5000, U.S.A.}

\emailAdd{cdshin@post.kek.jp}

\abstract
{
    This manuscript describes an innovative method to tag muons using the baseline information of the Flash ADC (FADC) waveform of PMTs in the JSNS$^{2}$ (J-PARC Sterile Neutrino Search at J-PARC Spallation Neutron Source) experiment. The experiment is designed to search for evidence of sterile neutrinos, and a reliable method for muon tagging is an essential component for background rejection because the detector is located above ground, on the 3rd floor of the J-PARC Material and Life Science Experimental Facility (MLF). Cosmogenic muons that stop within the detector volume and produce a Michel electron are a particularly important background that must be rejected for our sterile neutrino search.  Utilizing this innovative method, more than 99.8\% of Michel electrons can be rejected even without using information from the detector's veto region PMTs. This technique can be employed by any experiments which uses a similar detector configuration.
}


\begin{document}
    \maketitle
    \flushbottom
    
    \section{Introduction}
    \label{sec:intro}
    
    The JSNS$^2$ experiment~\cite{JSNS2_proposal, JSNS2_TDR}, proposed in 2013,
    searches for sterile neutrinos, which could explain anomalous observations made over more than 20 years~\cite{CITE:LSND, CITE:BEST, CITE:MiniBooNE2018, CITE:Reactor}. 
    The experiment consists of a liquid scintillator detector which 
    contains 17 tonnes of Gadolinium loaded liquid scintillator (Gd-LS) to 
    search for $\bar{\nu}_{\mu} \to \bar{\nu}_{e}$ oscillations with 24~m baseline.
    The detector is located on the 3rd floor of J-PARC's Material and Life Science Experimental Facility (MLF).
    An intense $\bar{\nu}_{\mu}$ flux is created from the interaction of the MLF's 3 GeV proton beam on the facilities liquid mercury target;
    the neutrinos of interest for the sterile neutrino search are produced via muon decay-at-rest ($\mu$DAR).
    The 3 GeV proton beam is delivered by J-PARC's Rapid Cycle Synchrotron
    (RCS) with a precise timing structure --- two proton bunches separated by
    600~ns with a 25~Hz repetition rate.
    This timing structure is essential to separate sterile neutrino signal events from background events.

    The liquid scintillator is ideally suited to detect the $\bar{\nu}_{e}$ since 
    the free protons inside the liquid scintillator can induce the Inverse-Beta-Decay (IBD)
    reaction: $\bar{\nu}_{e} + p \to e^{+} + n$. This reaction provides two coincident 
    signals, the prompt signal from the positron and the delayed signal from the gammas
    produced by the n-Gd capture reaction.
    Data taking started in 2020, and as of June 2024 data from 
    4.9$\times 10^{22}$ Proton-On-Target (POT) has been acquired.
    

    The detector consists of a target, a gamma catcher, and a cosmic ray veto layers.
    The target layer contains 17~tonnes of Gd-LS, while other layers contain 31~tonnes of pure
    liquid scintillator (pure-LS, no Gd). The latter two layers have a nominal thickness of 25 cm, and the target Gd-LS is contained by a cylindrical acrylic vessel that is
    2.5~m in height and 3.2~m in diameter. The veto layer is utilized to detect incoming particles 
    from outside of the detector, particularly cosmogenic muons.
    Ninety-six 10-inch R7081 Photo-Multiplier-Tubes 
    (PMTs)~\cite{CITE:Hamamatsu, CITE:Matsubara} located in the catcher region surround the 
    acrylic vessel to detect neutrino events, and twenty-four PMTs are located within the veto 
    region. 
    PMT signals are recorded by 28 CAEN v1721 500 MS/s FADC modules.
    Using Front-End-Electronics (FEEs) modules donated by the Double-Chooz experiment,
    JSNS$^2$ data acquisition system has effectively four volts dynamic range and 
    16 bits resolutions with a 500~MHz sampling rate~\cite{CITE:DAQ, CITE:NIM}.
    Further details on the JSNS$^2$ detector can be found in reference~\cite{CITE:NIM}.

    \section{Backgrounds induced by muons}

\begin{figure}[htbp]
       \centering
       \includegraphics[width=.58\textwidth]{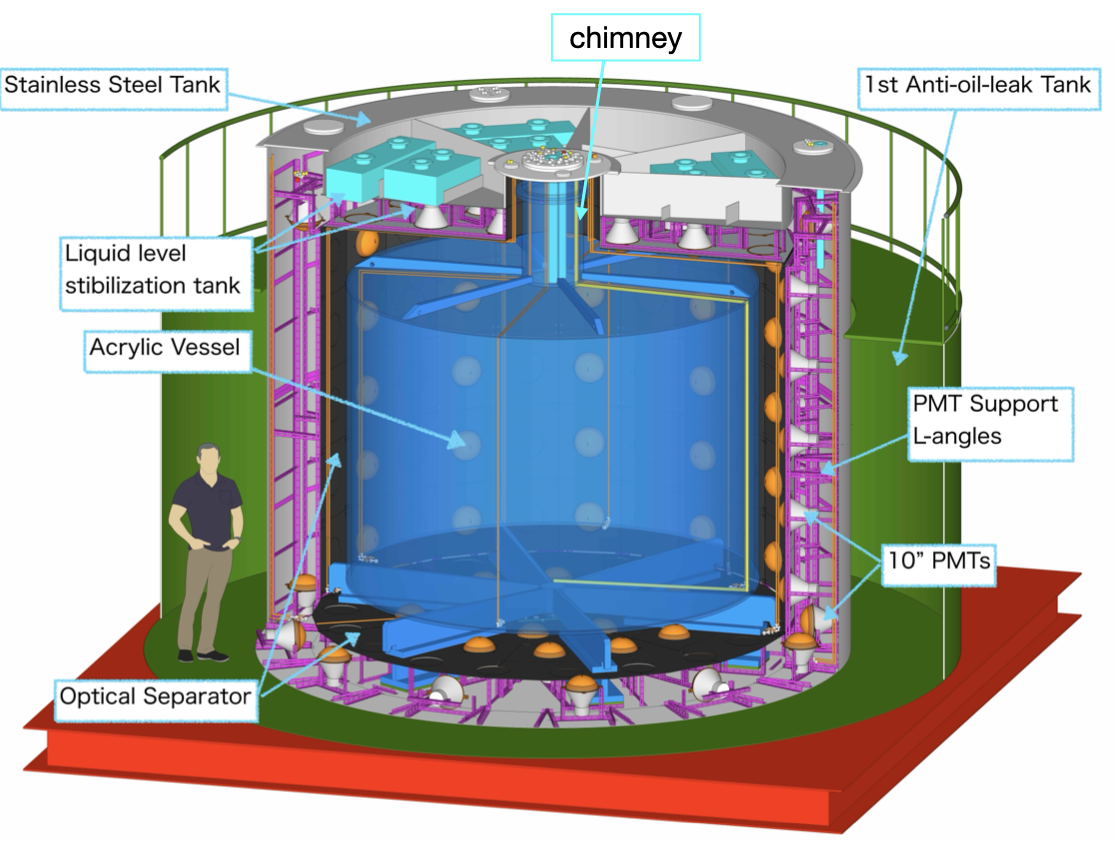}
       \caption{Structure of the JSNS$^{2}$ detector}
       \label{fig:Detector1}
   \end{figure}

    The JSNS$^2$ detector is located above ground, thus cosmic ray induced backgrounds 
    are significant. This background consists of cosmic ray muons at about 2 kHz, with a
    stopping muon rate of $\sim100$ Hz.
    Using only information from the detector veto region cosmogenic muons can be rejected
    with greater than 99 $\%$ efficiency, however some muons fail to create an observable signal in the veto
    layer if they pass the acrylic vessel's centrally located chimney region as shown in a schematic drawing of the detector (Fig.~\ref{fig:Detector1}).

    Another cause for untagged muons in the detector veto originates from the JSNS$^2$'s trigger dead time.
    The trigger dead time is created to avoid overlapping FADC readout windows.

    Untagged muons that stop within the detector volume and
    create a Michel electron is a significant source of
    backgrounds for JSNS$^2$~\cite{CITE:Neutrino2022}. High energy muons can also cause
    deep inelastic scattering events, which 
    contain Michel electrons and neutrons in sequence. 
    To tag and reject these backgrounds from untagged muons a new technique,
    described in the next section, was implemented for JSNS$^2$.

    \section{A new method to tag muons}
    
\begin{figure}[htbp]
        \centering
        \includegraphics[width=.7\textwidth]{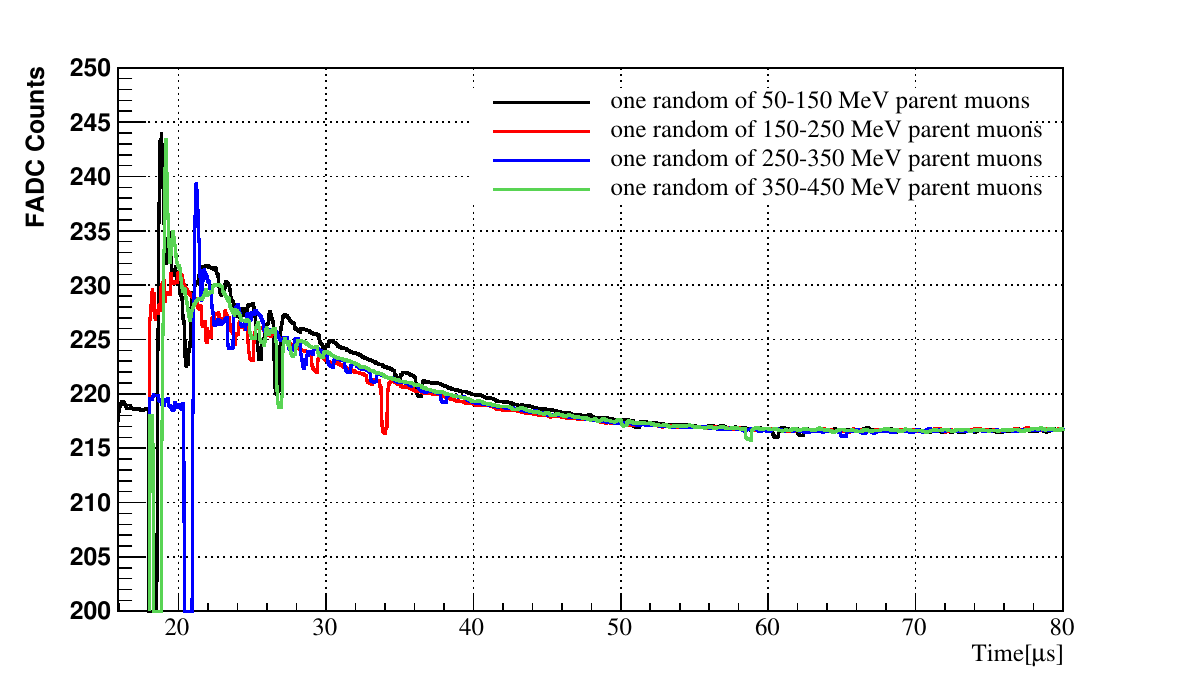}
        \caption{The typical PMT waveform baseline shift after the muons tagged by the detector veto region using the special calibration runs with a 125 \textmu{}s FADC gate and the beam schedule timing trigger.}
        \label{fig:baseline_shift0}
    \end{figure}

    In order to reduce the backgrounds associated with the veto-untagged muons,
    we developed an innovative technique. When a 10-inch PMT in JSNS$^2$ receives a highly luminous scintillation light signal from an energetic particle such as a muon, the baseline of the waveform after the particle is shifted.
Fig.~\ref{fig:baseline_shift0} shows the typical baseline shift after the muons tagged by the detector veto region using the special calibration runs with a 125 \textmu{}s FADC gate and the beam scheduled timing trigger~\cite{CITE:DAQ, CITE:NIM}. Different colors show the energy dependence of muons which were picked up randomly. The baseline shift typically continues $\sim$25 \textmu{}s in JSNS$^2$, independent on their energies. This baseline shift can be used to tag muons clearly.

\begin{figure}[htbp]
        \centering
        \includegraphics[width=.95\textwidth]{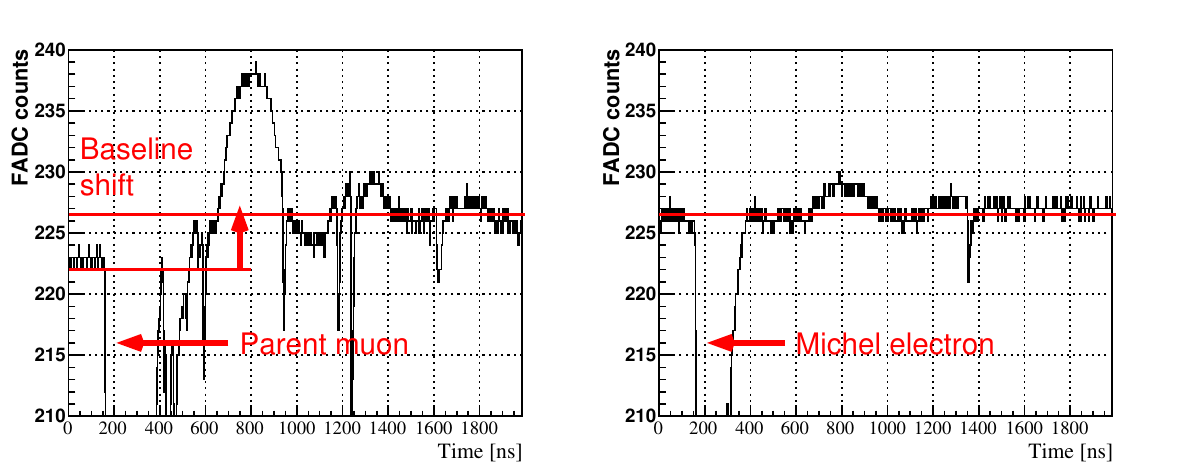}
        \caption{The typical PMT waveform baseline shift.}
        \label{fig:baseline_shift}
    \end{figure}

    Fig.~\ref{fig:baseline_shift} shows the typical baseline shift for an individual PMT baseline for
    the parent muon and for the following Michel electron during the normal data taking.
    JSNS$^2$ uses a 2 \textmu{}s of FADC time gate during the normal data taking. 

    The baseline shift is due to the charge recovery process after a large signal for 
    a coupling capacitor in the splitter circuit. 
    The splitter circuit separates the PMT's light signals from the high-voltage supply. 
    This phenomenon is widely known, and reference~\cite{CITE:baseline} describes it in detail.
    In JSNS$^2$'s case, the PMT {bleeder} circuits~\cite{CITE:Matsubara},
    the splitters~\cite{CITE:DCsplitter}, Front-End-Electronincs ( FEE~\cite{CITE:DCsplitter}), 
    and the lengths of the PMT cables (RG303U, typically 22 meters) and those of the signal cables (RG58C/U, typically 15 
    meters) affect the time constant. Note, the splitter and FEE circuits were 
    donated to JSNS$^2$ from the Double-Chooz experiment.


    
    We use the number of PMTs exhibiting this kind of baseline shift to tag muon events.
    Hereafter, the method to tag muons and the corresponding efficiencies are described.

    \subsection{Control sample for this study}
    
\begin{figure}[htbp]
        \centering
        \includegraphics[width=.45\textwidth]{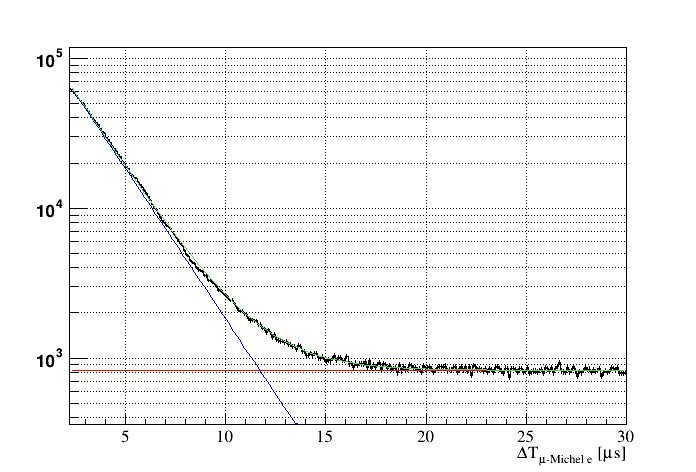} \includegraphics[width=.45\textwidth]{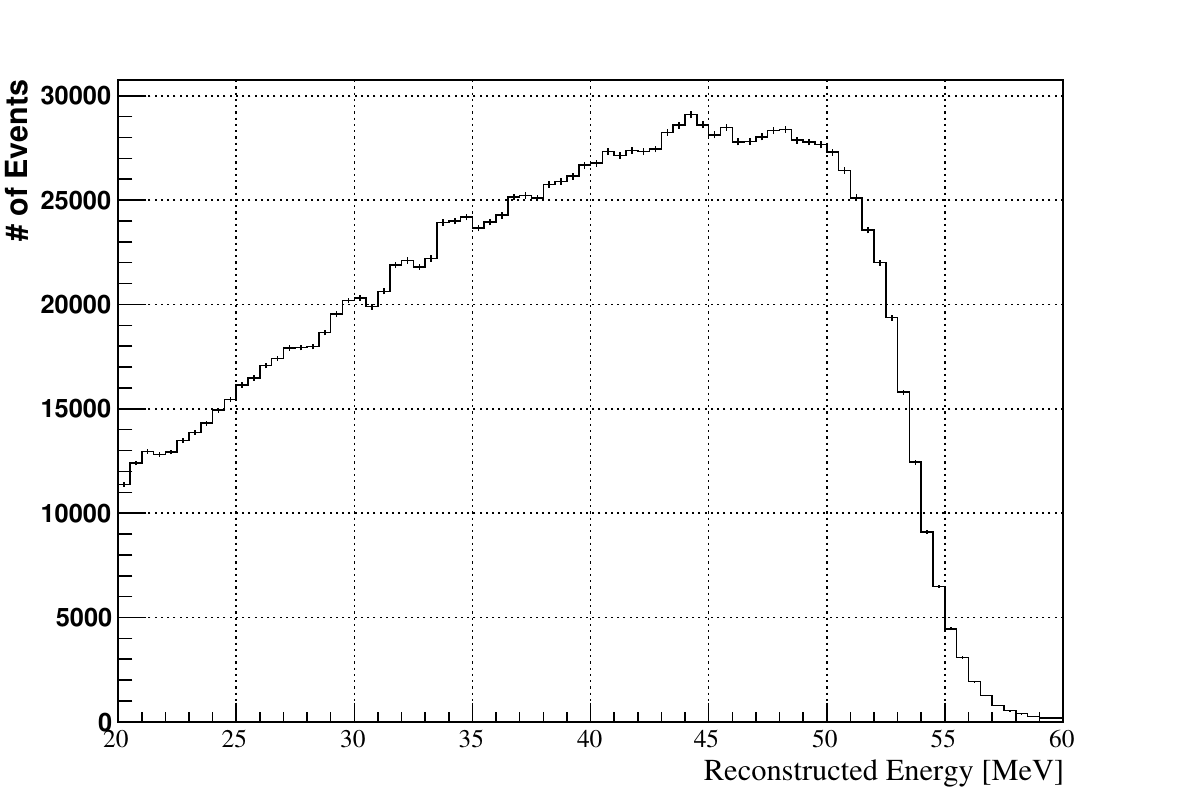}
        \caption{The left panel shows the timing difference between the parent muons and the Michel electrons.
        The right panel shows the energy spectra of the Michel electrons.}
        \label{fig:control_sample}
    \end{figure}

    To create a control sample of parent muons and their decay Michel electrons
    for this study, delayed events are selected with a reconstructed energy range of
    20-60 MeV within 10 \textmu{}s from the parent muons as shown
    in Fig.~\ref{fig:control_sample}.
    Parent muons here are defined as events where more than 100 photo-electrons (p.e.) are observed in the veto region PMTs.
    The timing difference between the parent muons and the Michel electrons is shown in the left panel.
    The shown fit function is composed of a constant term and an exponential decay term as shown in Eq.~\ref{eq:fit_func}.

\begin{equation}
f(x) = p_0 \cdot \exp(-x / p_1) + p_2
\label{eq:fit_func}
\end{equation}

The best fit parameters for this function are:
\begin{align*}
p_0 &= (1.85 \pm 0.06) \times 10^6  \\
p_1 &= 2.16 \pm 0.03~\mu\mathrm{s} \\
p_2 &= (8.27 \pm 0.03) \times 10^2 
\end{align*}

     The right panel of Fig.~\ref{fig:control_sample} shows the energy spectrum of the Michel electrons. 
    This control sample is used to evaluate the tagging efficiency and 
    the fake rate of the signal sample.

    \subsection{Definition of the nominal FADC counts of the waveform baseline}
    
\begin{figure}[htbp]
        \centering
        \includegraphics[width=.62\textwidth]{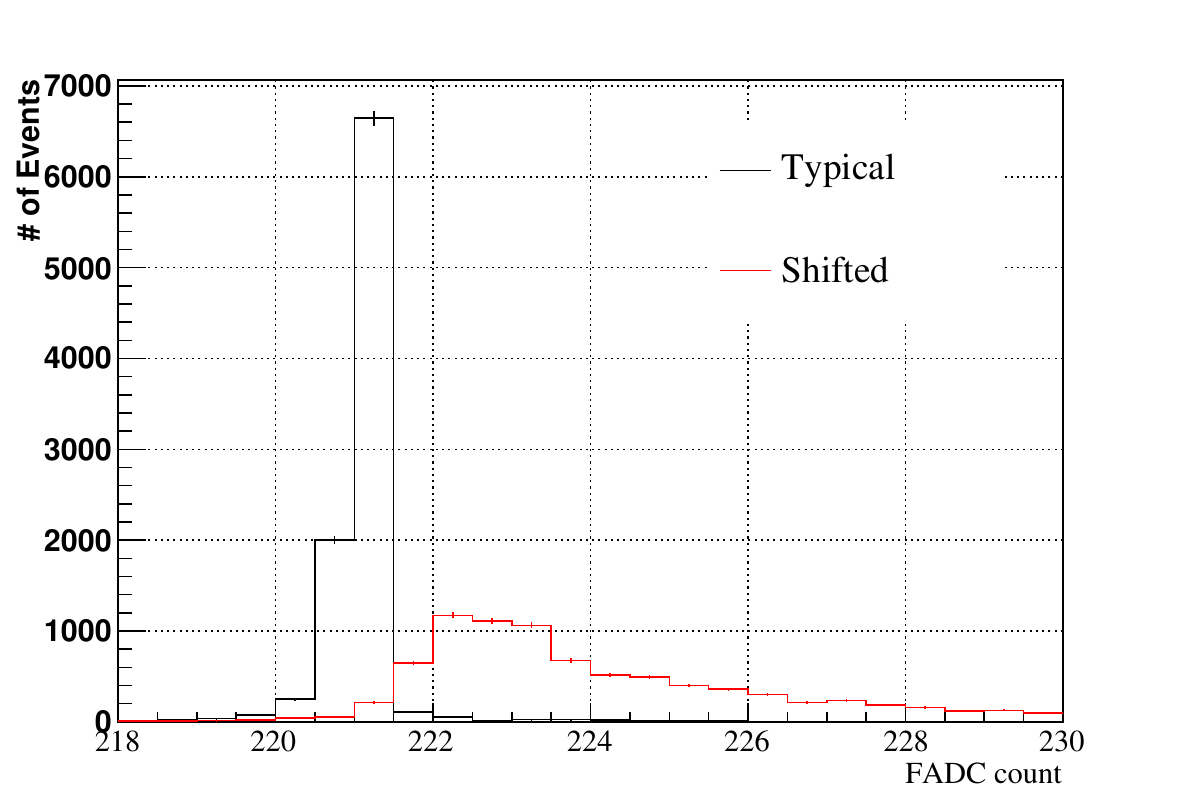}
        \caption{Typical waveform baseline FADC count (black) and shifted waveform baseline FADC count (red) of one PMT. }
        \label{fig:pedestal}
    \end{figure}

    Each PMT has different waveform baselines and small fluctuations 
    even without muon events in JSNS$^2$, therefore we first define the nominal FADC count of the waveform 
    baseline for each PMT.
    The nominal FADC count baseline value is determined from the beginning of the
    waveform for parent muon events,
    assuming there are no previous events.
    Fig.~\ref{fig:pedestal} (black) shows the determined baseline value distribution
    for a typical PMT.
    As part of the charge computing process for each event the pedestal (baseline) value of the FADC 
    waveform is determined from a 60 ns window within the 2000 ns FADC readout,
    the window is chosen such that the variance of samples within the window
    is minimal.

    A Gaussian approximation cannot describe the distribution of baseline values, thus
    we define a nominal baseline FADC count range as within typically $\pm 2$ FADC counts around the mean, which contains 98.2 \% of events. The mean value of ``nominal” baseline are different in each PMT, hence the range for nominal baseline FADC count ranges are calculated individually containing 98.2 \%


    The distribution of shifted baselines for Michel electron events, observed by the same PMT,
    are shown as Fig.~\ref{fig:pedestal} (black);
    the two distributions are clearly distinct.

    \subsection{Number of PMTs which have shifted baselines}
    
\begin{figure}[htpb]
        \centering
        \includegraphics[width=.7\textwidth]{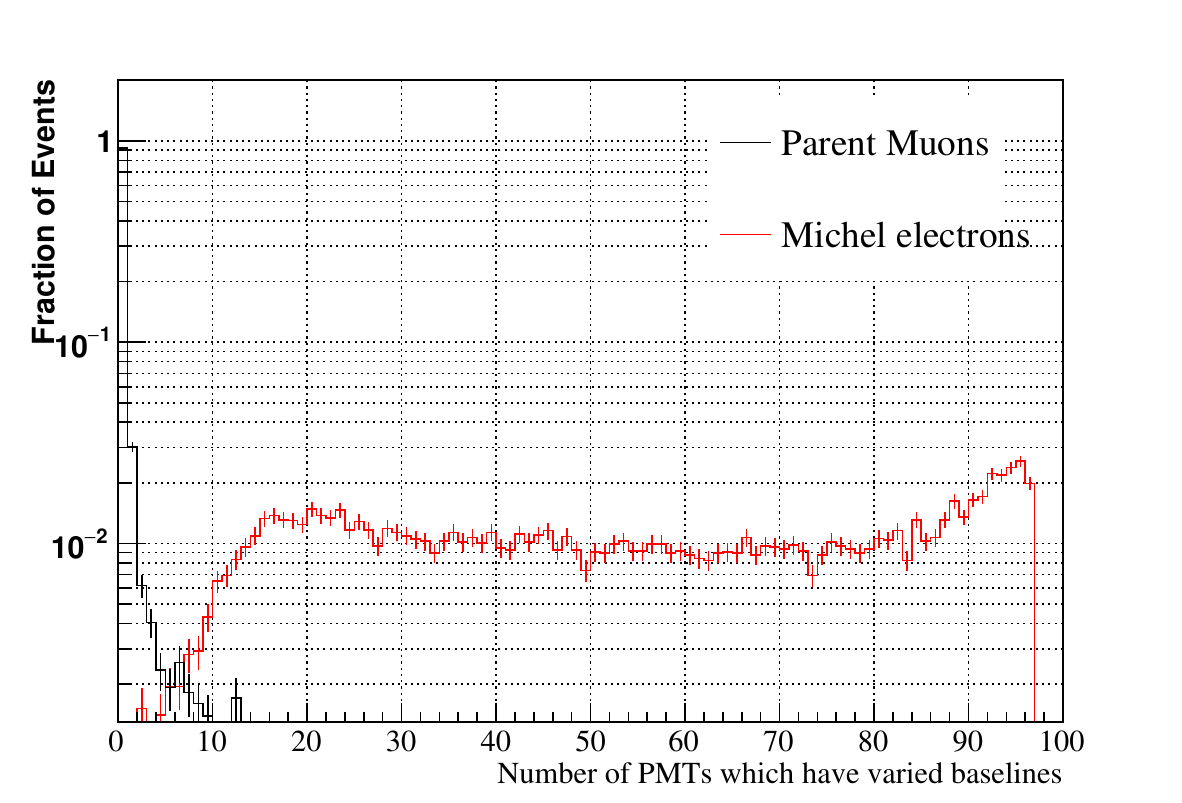}
        \caption{Number of PMTs which have shifted baselines. Black: parent muons, red: Michel electrons.}
        \label{fig:NPMT}
    \end{figure}

    Shifted waveform baselines are expected to appear not only in one PMT, but should appear in 
    multiple PMTs. Thus, thenumber of PMTs which have 
    shifted baselines ($N_{PMT}$)  are examined using the control sample.
    A PMT is counted in $N_{PMT}$ if its baseline value falls outside of its estimated
    98.2 \% range.
    Parent muon events are expected to have nominal waveform baselines because they are 
    defined by pedestal prior to muon's arrivals, while the 
    following decay electrons will have some number shifted baselines.
    Fig.~\ref{fig:NPMT} shows the number of PMTs which have shifted baselines. 
    The black histogram shows the distributions from parent muon events, while the red
    corresponds to Michel electrons events.
    99.86$\pm$0.05 \% of Michel electron events have N$_{PMT}$ $\ge$ $2$,
    however only 4.9$\pm$0.2 \% of parent muons are mistakenly tagged by the same criteria.
    Note that this number for parent muons indicates the presence of prior activity. Given a typical baseline shift duration of $\sim$25 \textmu{}s and a cosmic muon rate of $\sim$2 kHz, a simple estimation yields an expected mis-tagging rate of approximately 5 \%, which is consistent with the observed rate in the control sample.  
    Thus it can be widely used for other event samples, such as IBD events in the sterile neutrino search.
    Due to the JSNS$^2$ trigger scheme, we have at least $\sim$5ms deadtime before beam timing.  Parent muons in our control sample may themselves have unrecorded or untagged muon events occurring just before the observed muon event, resulting in a source of inefficiency which is similar condition with real sterile neutrino search.


    \subsection{Efficiencies}

\begin{figure}[htbp]
        \centering
        \begin{subfigure}[b]{0.47\textwidth}
            \includegraphics[width=1.\textwidth]{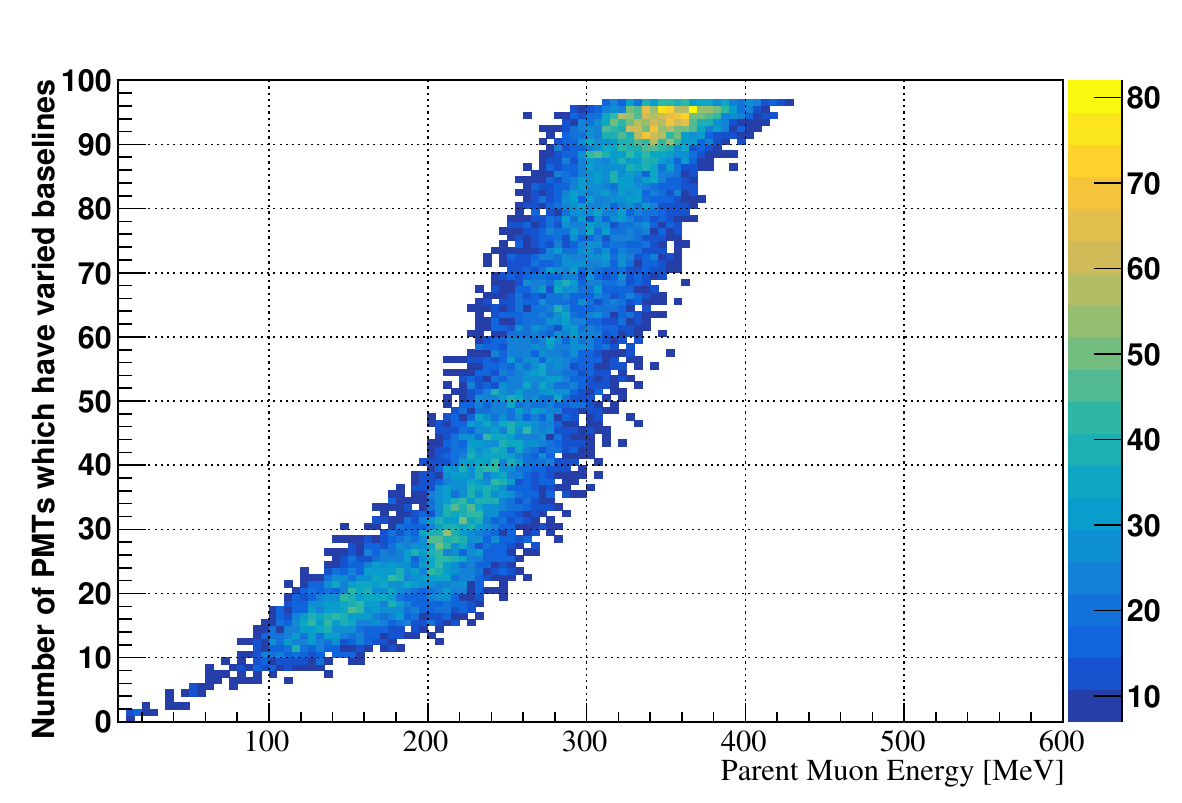}
            \caption{ Energy dependence }
            \qquad
        \end{subfigure}
        \begin{subfigure}[b]{0.47\textwidth}
            \includegraphics[width=1.\textwidth]{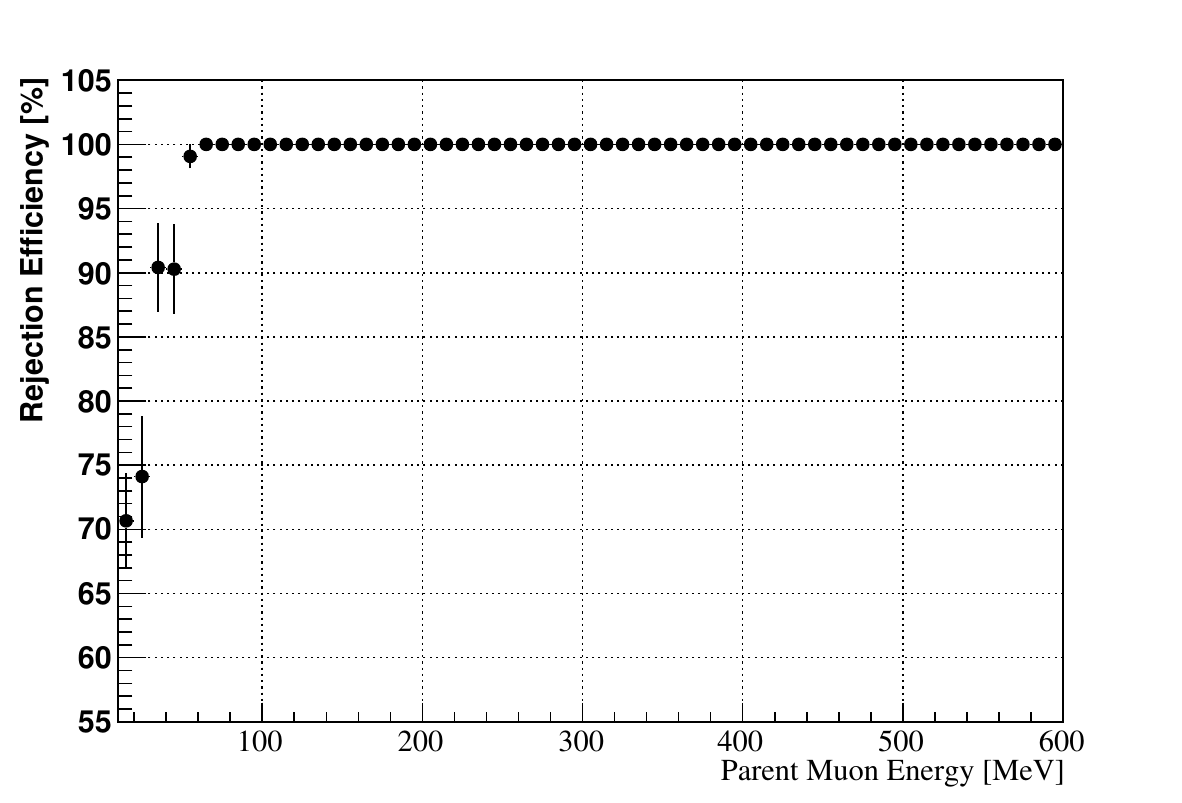}
            \caption{ Rejection efficiency }
            \qquad
        \end{subfigure}
        \caption{Left : Parent muon energies (horizontal) vs number of PMTs which have shifted 
        waveform baselines (vertical). Right : Rejection efficiency as a function of energy of the parent muons.}
        \label{fig:Edep}
    \end{figure}

    This muon tagging method depends on the energy of the parent muon, and the time difference between
    parent muon and the Michel electron, thus the efficiency is a function of these variables;
    the time and energy depedence can be evaluated within the control sample.
    The left plot of Fig.~\ref{fig:Edep} shows the two dimensional histogram of reconstructed parent muon
    energy and $N_{PMT}$. A clear correlation is observed.
    The right plot shows the tagging 
    efficiency ($N_{PMT} < 2$) as a function of parent muon energy.
    Parent muons which have more than 70 MeV, which is the dominant majority
    of events, have a 100 \% tagging efficiency within our control sample.
    
\begin{figure}[htbp]
        \centering
        \begin{subfigure}[b]{0.47\textwidth}
            \includegraphics[width=1.\textwidth]{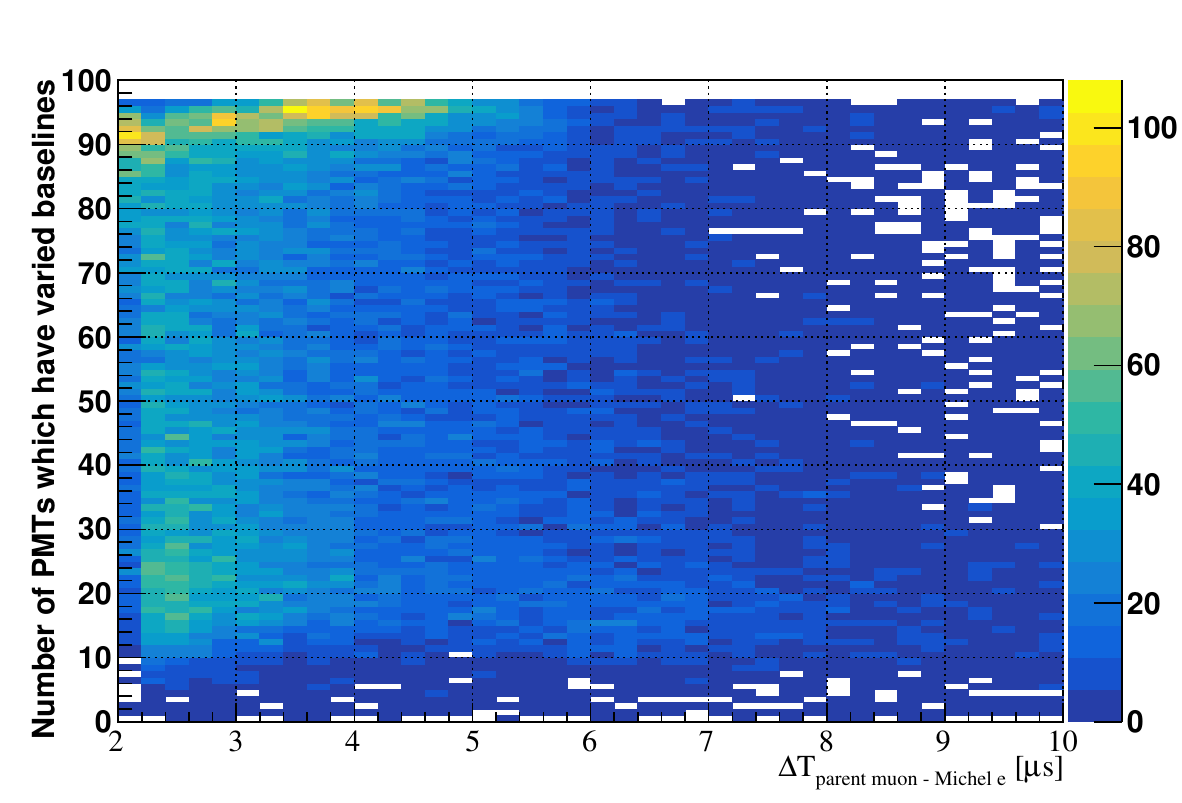}
            \caption{ $\Delta T$ dependence }
            \qquad
        \end{subfigure}
        \begin{subfigure}[b]{0.47\textwidth}
            \includegraphics[width=1.\textwidth]{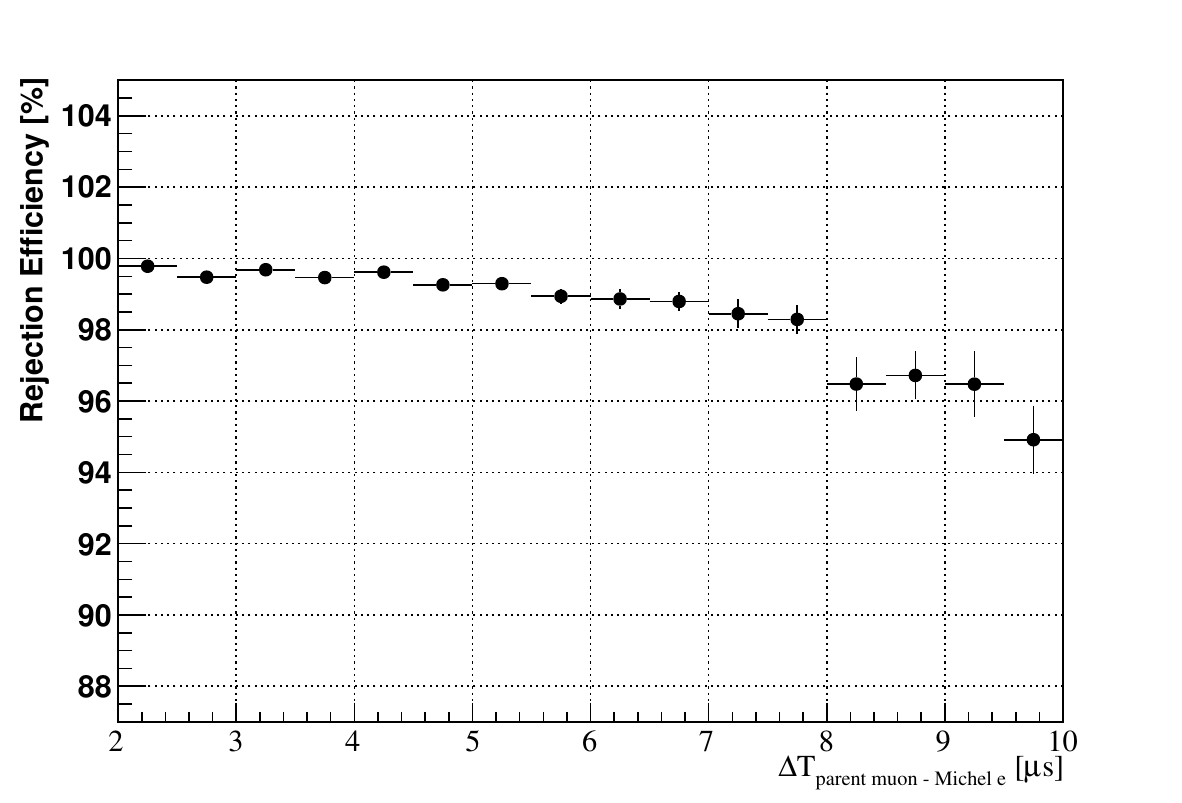}
            \caption{ Rejection efficiency }
            \qquad
        \end{subfigure}
        \caption{Left : $\Delta T$ (\textmu{}s : horizontal) vs number of PMTs which have shifted 
        waveform baselines (vertical). Right : Rejection efficiency as a function of $\Delta T$.}
        \label{fig:DTdep}
    \end{figure}

    The left plot of Fig.~\ref{fig:DTdep} shows the two dimensional histogram of the time difference ($\Delta T$) between
    parent muons and Michel electrons vs 
    $N_{PMT}$.
    No clear correlation is seen in this case.
    The right plot shows the tagging efficiency ($N_{PMT} < 2$) as a function of the $\Delta T$.
    The efficiency is reduced for events with $\Delta T$ greater than approximately 6~\textmu{}s,
    however, even for those events the tagging efficiency is at least 94 \%.
    For events with $\Delta T < 2\mu$s the parent muon and Michel electron will both
    be recorded within the same FADC readout window. For that case, it becomes impossible to distinguish between a nominal and a shifted baseline since both the parent muon and the Michel electron share the same one pedestal baseline value with the minimal variance corresponding to that before parent muons. For this reason, these events are rejected in JSNS$^2$, with only $\sim0.4$ \% of inefficiency.
    Quantitatively, this inefficiency is calculated from the cosmic muon rate in JSNS$^2$ ($\sim2$ kHz) and the FADC time gate (2 \textmu{}s).

    \section{Summary}
    
    An  innovative technique to tag muons using the baseline information of the FADC waveform
    was developed by the JSNS$^2$ experiment. With this method, more than 99.8 \% muons 
    can be tagged while less than 5 \% of physics events are lost.
    This technique can be used to tag any hidden muons that are missed by the detector veto 
    region. This methodology could be applied by other experiments which 
    suffer from background muons, especially if Michel electrons 
    are associated with parent muons.

    \acknowledgments
We deeply thank J-PARC for their continued support, 
especially for the MLF and the accelerator groups that provide 
an excellent environment for this experiment.
We acknowledge the support of the Ministry of Education, Culture, Sports, Science, and Technology (MEXT) and the JSPS grants-in-aid: 16H06344, 16H03967, 23K13133, 
24K17074 and 20H05624, Japan. This work is also supported by the National Research Foundation of Korea (NRF): 2016R1A5A1004684, 17K1A3A7A09015973, 017K1A3A7A09016426, 2019R1A2C3004955, 2016R1D1A3B02010606, 017R1A2B4011200, 2018R1D1A1B07050425, 2020K1A3A7A09080133, 020K1A3A7A09080114, 2020R1I1A3066835, 2021R1A2C1013661, NRF-2021R1C1C2003615, 2021R1A6A1A03043957, 2022R1A5A1030700, RS-2023-00212787 and RS-2024-00416839. Our work has also been supported by a fund from the BK21 of the NRF. The University of Michigan gratefully acknowledges the support of the Heising-Simons Foundation. This work conducted at Brookhaven National Laboratory was supported by the U.S. Department of Energy under Contract DE-AC02-98CH10886. The work of the University of Sussex is supported by the Royal Society grant no. IESnR3n170385. We also thank the Daya Bay Collaboration for providing the Gd-LS, the RENO collaboration for providing the LS and PMTs, CIEMAT for providing the splitters, Drexel University (courtesy of Prof. Chunk) for providing the FEE circuits and Tokyo Inst. Tech for providing FADC boards.

    
    

\end{document}